\documentclass[onecolumn,floatfix,prb,amsmath,amssymb,showpacs]{revtex4}

\usepackage{amsmath}
\usepackage{amssymb}
\usepackage{epsfig}
\usepackage{latexsym}
\usepackage{multirow}
\usepackage{natbib}
\usepackage{psfrag}
\usepackage{hyperref}

\hypersetup{
  colorlinks = true,
  citecolor  = blue,
  linkcolor  = blue,
  urlcolor   = blue
}

\newlength{\figurewidth}
\begin{document}

\title{
  Optical Conductivity in a Two--Band Superconductor: Pb
}

\date{\today}

\author{N.~Bock}
\email{nbock@lanl.gov}

\affiliation{
  Theoretical Division, Los Alamos National Laboratory, Los Alamos, New Mexico
  87545
}

\author{D.~Coffey}

\affiliation{
  Dept. of Physics, Buffalo State College, Buffalo, New York 14222
}

\date{\today}

\pacs{74.25.Nf,74.25.Gz, 74.25.Jb}

\begin{abstract}

  We demonstrate the effect of bandstructure on the superconducting properties
  of Pb by calculating the strong-coupling features in the optical conductivity,
  $\sigma(\omega)$, due to the electron-phonon interaction.   The importance of
  momentum dependence in the calculation of the properties of superconductors
  has previously been raised for MgB$_2$\cite{Choi2002a, Choi2002b}. Pb
  resembles MgB$_2$ in that it is a two band superconductor  in which the bands'
  contributions to the Fermi surface have very different topologies.  We
  calculate $\sigma(\omega)$ by calculating a memory function\cite{Goetze72}
  which has been recently used to analyze $\sigma(\omega)$ of
  Bi$_2$Sr$_2$CaCu$_2$O$_{8+\delta}$\cite{Hwang2004}.  In our calculations the
  two components of the Fermi surface are described by parameterizations of de
  Haas--van Alphen data.  We use a phonon spectrum which is a fit to neutron
  scattering data.  By including the momentum dependence of the Fermi surface
  good agreement is found with the experimentally determined strong-coupling
  features which can be described by a broad peak at around 4.5~meV  and a
  narrower higher peak around 8~meV of equal height. The calculated features are
  found to be dominated by scattering between states within the third band.  By
  contrast scattering between states in the second band leads to strong-coupling
  features in which the height of the high energy peak is reduced by $\sim 50\%$
  compared to that of the low energy peak.  This result is similar to that in
  the conventional isotropic (momentum independent) treatment of
  superconductivity.  Our results show that it is important to use realistic
  models of the bandstructure and phonons, and to avoid using momentum averaged
  quantities, in calculations in order to get quantitatively accurate results.
  {\bf LA-UR-06-4303}

\end{abstract}

\maketitle

\section{Introduction}

\citet{Suhl1959} originally introduced a two-band model for superconductors as
an extension of the original BCS calculation in which each band has its own
superconducting gap.  This two-band model was proposed to explain fine structure
in the infrared absorption of superconductors with $s$ and $d$ electrons.  Many
years later \citet{Allen76} proposed a formalism to extend the Eliashberg
treatment of superconductivity to include multiband Fermi surfaces and Fermi
surfaces with complicated geometry by expanding order parameters, pairing
interactions and scattering mechanism in terms of Fermi surface harmonics.  This
formalism was employed by \citet{Golubov1997} to investigate the effect of
impurity scattering on the value of the critical temperature in  two-band model
superconductors.  \citet{Kresin1992} also discussed two band superconductivity
in the context of the high $T_{c}$ materials in which superconductivity in one
band is induced in a second band by interband scattering.  These investigations
were not extended to get quantitative results for specific materials.
Quantitative investigations of superconductivity in real multiband systems have
only been developed in recent years with the availability of the necessary
computational resources. Renewed interest in this area is driven by experiment.
In 2001 \citet{Nagamatsu2001} discovered that MgB${}_{2}$ becomes a
superconductor at temperatures below $T_{c} = 39$ K. This value of $T_{c}$ is
comparable to that of La$_{1-x}$Sr$_x$CuO${}_{4}$, the first of the cuprate
superconductors to be discovered. This discovery initiated a number of
theoretical and experimental studies on the nature of the superconducting state
in this material.  MgB${}_{2}$ shows a significant isotope effect, a tell--tale
sign of the involvement of phonons in the process\cite{Budko01}. Several
tunneling experiments found very different values for the superconducting
gap\cite{Rubio-Bollinger01, Schmidt01}, which led \citet{Liu01} to suggest that
the superconducting state of MgB${}_{2}$ exhibits multiple gaps.  First
principle calculations\cite{Kortus01} found essentially two distinct conduction
bands, and it is this two--band nature of the MgB${}_{2}$ system that is
responsible for two different gaps. The calculated Fermi surface was
subsequently found to be in agreement with angle resolved photoemission
(ARPES)\cite{Uchiyama2002} and de Haas--van Alphen\cite{Carrington2003}
measurements. \citet{Choi2002a, Choi2002b} calculated the gap function and
$T_{c}$ using the anisotropic Eliashberg formalism\cite{Allen82} which
incorporates the momentum dependence of the phonons, electronic bandstructure
and the electron-phonon interaction.  Choi et al.  used the {\it ab initio}
pseudopotential density-functional method to calculate these.  They pointed to
the strong variation in the electron-phonon interaction on and between the Fermi
surfaces of the two bands in which the interband scattering is much weaker than
intraband scattering because of the different symmetry of bands in
MgB$_2$.\cite{Choi2002a, Choi2002b} The strong dependence of the electron-phonon
coupling on Fermi surface states in MgB$_2$ leads to a wide range of values for
the superconducting gap.  Choi et al.  emphasized the importance of including
the momentum dependence of these quantities rather than using the isotropic
version of the Eliashberg equations for quantitatively accurate results.  Here
we show that there is also a strong dependence on bandstructure in the optical
conductivity of superconducting lead.  Pb is effectively a two band material in
which the electron-phonon interband scattering is comparable to the intra-band
scattering so that the superconducting gaps on the two bands are almost equal,
in contrast to MgB$_2$. $T_{c} \sim 7.9$ K, is also much lower in Pb than in
MgB$_2$.  In order to investigate the importance of including the different
character of the two bands which cross the Fermi level in Pb we calculate the
strong-coupling features which appear in optical conductivity.  

Optical conductivity can be thought of as having two contributions.  The first
one is associated with the anomalous skin effect, whereas the second is due to
``bulk'' properties of the metal.  The anomalous skin effect depends on the
nature  of the scattering of electrons from the surface and arises from the
breakdown of momentum conservation at the surface.  This contribution was
initially investigated theoretically by a number of authors in the normal
state\cite{Reuter48, Chambers52,
Pippard54:Advances_in_Electronics_and_Electron_Physics}. In superconductors it
was studied by Pippard using the phenomenological London
model\cite{Pippard54:Advances_in_Electronics_and_Electron_Physics} and by Mattis
and Bardeen on the basis of the BCS theory\cite{Mattis58}.  The second
contribution, the ``bulk'' part, comes from scattering of the incident photons
on phonons, impurities, or other forms of scattering centers, within a distance
comparable to the skin depth from the surface. We will focus on the Holstein
process, the simultaneous absorption and emission of phonons by an electron, and
ignore impurity and electron-electron scattering in our calculations using a
memory function approach. This has previously been investigated by
Allen\cite{Allen71}.

\citet{Farnworth74, Farnworth76} studied the strong--coupling effects associated
with the electron--phonon interaction by infrared absorption. They measured the
frequency dependence of the light emitted from a cavity whose walls contain Pb.
They analyzed their data by taking the derivative of the ratio of the signal
from the cavity in the normal and the superconducting states which leads to a
frequency dependence given by the difference of the derivatives of the
absorptivity in the superconducting and normal states, $A_S( \omega)$ and $A_N(
\omega)$.  This procedure  has the effect of subtracting out the anomalous skin
effect contribution and of enhancing the difference between the normal and the
superconducting states.  This results in a transport version of the $\alpha^{2}
F$ function which arises in the Eliashberg formalism. 

\begin{figure}
 \epsfig{file=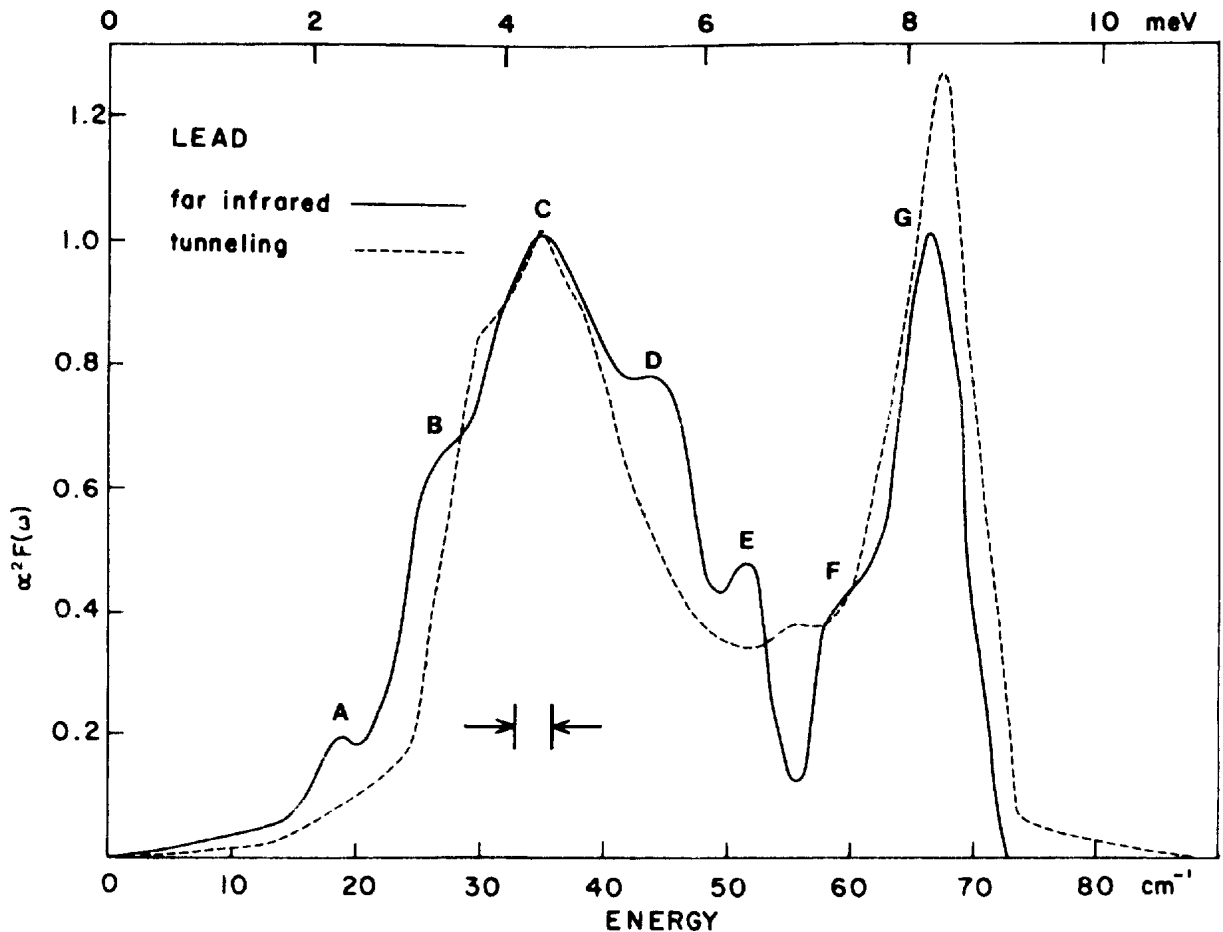,angle=0,width=\figurewidth}
 \caption{
    \label{fig_2_farnworth76}
    Reproduced from reference 24 with permission: Comparison of $\alpha^{2} F$
    numerically inverted from infrared absorption (solid line) and tunneling
    (dashed line)
  }
\end{figure}

Farnworth and Timusk's data, shown in Fig.~\ref{fig_2_farnworth76}, exhibits
features associated with the phonon density of states at B, C, D, E, F, and G.
In addition the data has a feature, A, at $\omega = 4 \, \Delta_{0}$, where
$\Delta_{0}$ is the value of the superconducting gap.  This feature could be
thought of arising either from the  phonon--mediated interaction between the
superconducting quasi--particles or as an effect due to the change in the phonon
self-energy in the superconducting state.  The leading contribution to this
feature is then fourth order in the strength of the electron-phonon coupling.
The data of Farnworth and Timusk on thin films reveals a splitting of this
feature into two slightly different but distinct energies indicating that we are
dealing with two conduction bands in Pb\cite{Farnworth74}.

We use the memory function formalism which was developed by \citet{Goetze72} to
calculate the contributions to $\sigma (\omega)$ due to scattering events from
phonons, impurities, and spin fluctuations. The assumption of this approach is
that the memory function, $M (\omega)$, can be expressed as a power series in
the electron--phonon interaction strength, which makes this approach in
principle a weak coupling approach.  The advantage of this approach, compared to
summing ladder diagrams, is that it is much easier to include the details of the
Fermi surface which leads to improved quantitative agreement with experiment for
Pb. Previously \citet{Tomlinson1977} have investigated the temperature
dependence of the resistivity using the same experimental data as we use for Pb
with different approximate solutions to the Boltzmann equation.
\citet{Leung1977} have also taken a similar approach for the resistivity of
aluminum. 

Recently \citet{Hwang2004} determined $M (\omega)$, referring to it as the
``optical single-particle self-energy'', in their analysis of $\sigma(\omega)$
in Bi$_2$Sr$_2$CaCu$_2$O$_{8+\delta}$ and compared it with a single-particle
self-energy derived from ARPES data.  The interpretation of $\sigma(\omega)$
data in the high $T_{c}$ materials, as well as other types of data, is
considerably more difficult than in conventional electron-phonon
superconductors. In high $T_{c}$ superconductors strong correlation effects are
evident in phenomena, such as the pseudogap, and are thought to be responsible
for the high value of $T_{c}$. These correlation effects and the absence of a
consensus on a suitable microscopic model make inversion of data and analysis
based on  weak coupling treatments problematic. \citet{Schachinger2006} have
recently investigated the inversion of $\sigma(\omega)$ data for conventional
and high $T_{c}$ superconductors using singular value decomposition and maximum
entropy techniques. 

In our calculations we use fits to the experimentally determined Fermi surface
determined from de Haas--van Alphen data and phonons from neutron scattering
data, rather than calculate these as Choi et al. did  in the case of MgB$_2$.
In Pb there are two $6s$ and two $6p$ electrons per atom leading to four bands.
De Haas--van Alphen measurements and a subsequent analysis of these by
\citet{Anderson65} show that in Pb the lowest lying band is completely inside
the Fermi surface, and that the second and third bands both cut across the Fermi
energy, whereas the fourth band lies above the Fermi energy.  The component of
the Fermi surface due to the second band has the topology of a sphere while that
due to the third band is multiply connected and has the form of  cylinders along
each edge of the first Brillouin zone.  The two band nature of Pb is due to the
second and third bands. The phonons are described by Cowley's
fit\cite{Cowley74a} to neutron scattering data.  

In sections \ref{sec_electron} and \ref{sec_electron_phonon_model} we discuss
our treatment of the charge carriers and the electron-phonon coupling.  In
section \ref{sec_results_optical_conductivity} we derive an expression from the
electron-phonon contribution to the memory function in the superconducting state
and an expression for the absorptivity in terms of the memory function.  After
calculating the derivative of the difference of the calculated absorptivities in
the normal and superconducting states we compare the result with the data of
\citet{Farnworth74, Farnworth76}.  The comparison demonstrates the importance of
scattering involving the third band and resolves the discrepancy between the
experimentally determined $\alpha^{2} F$ and that calculated ignoring momentum
conservation and approximating the Fermi surface with a sphere\cite{Allen71}.

\section{Two-Band Character}
\label{sec_electron}

\begin{figure}
  \psfig{file=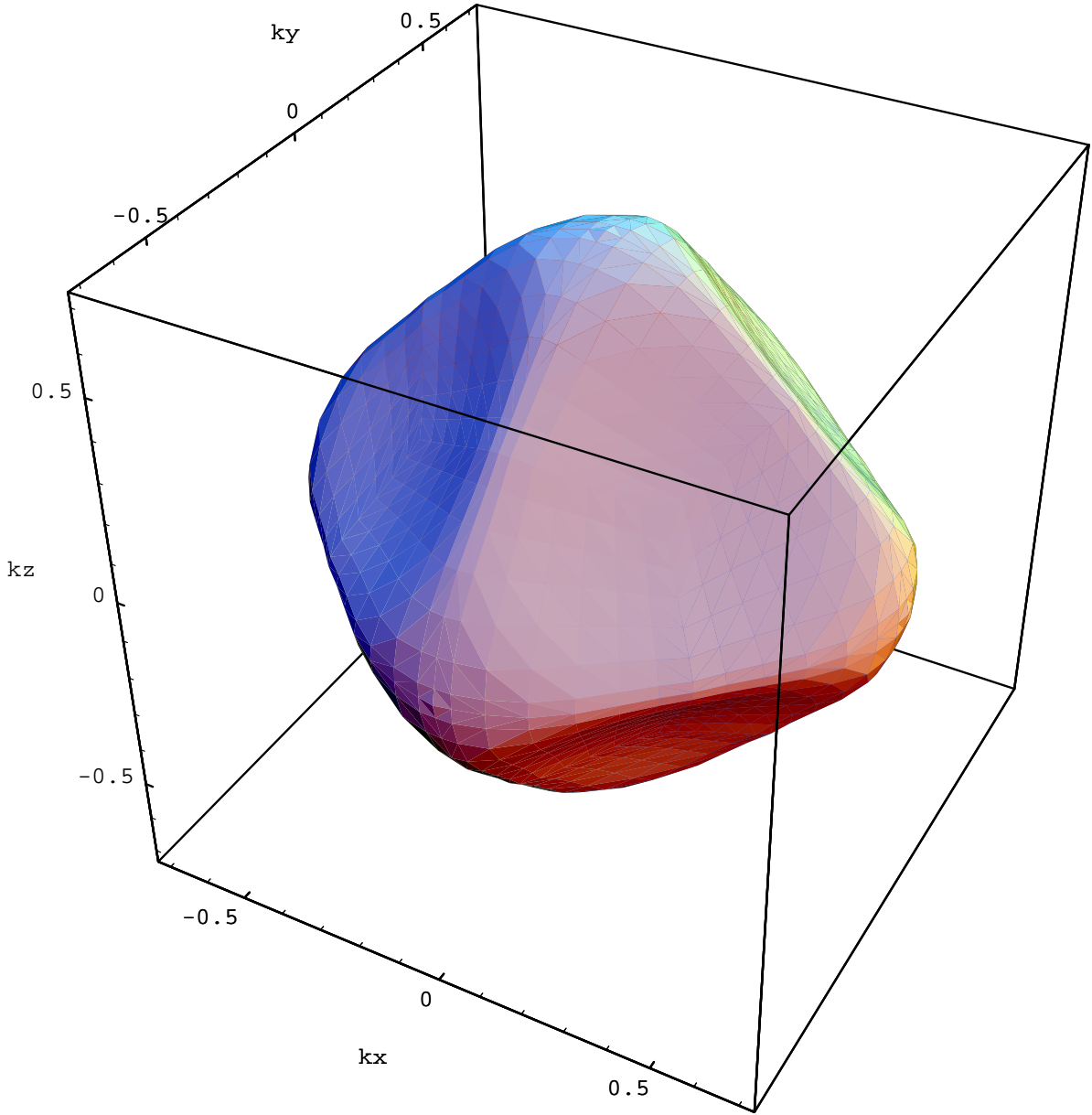,width=\figurewidth}
  \caption{
    \label{fig_Pb_2nd_band}
    (Color online) Second band in Pb calculated using the matrix by
    \citet{Anderson65}.
  }
\end{figure}

\begin{figure}
  \psfig{file=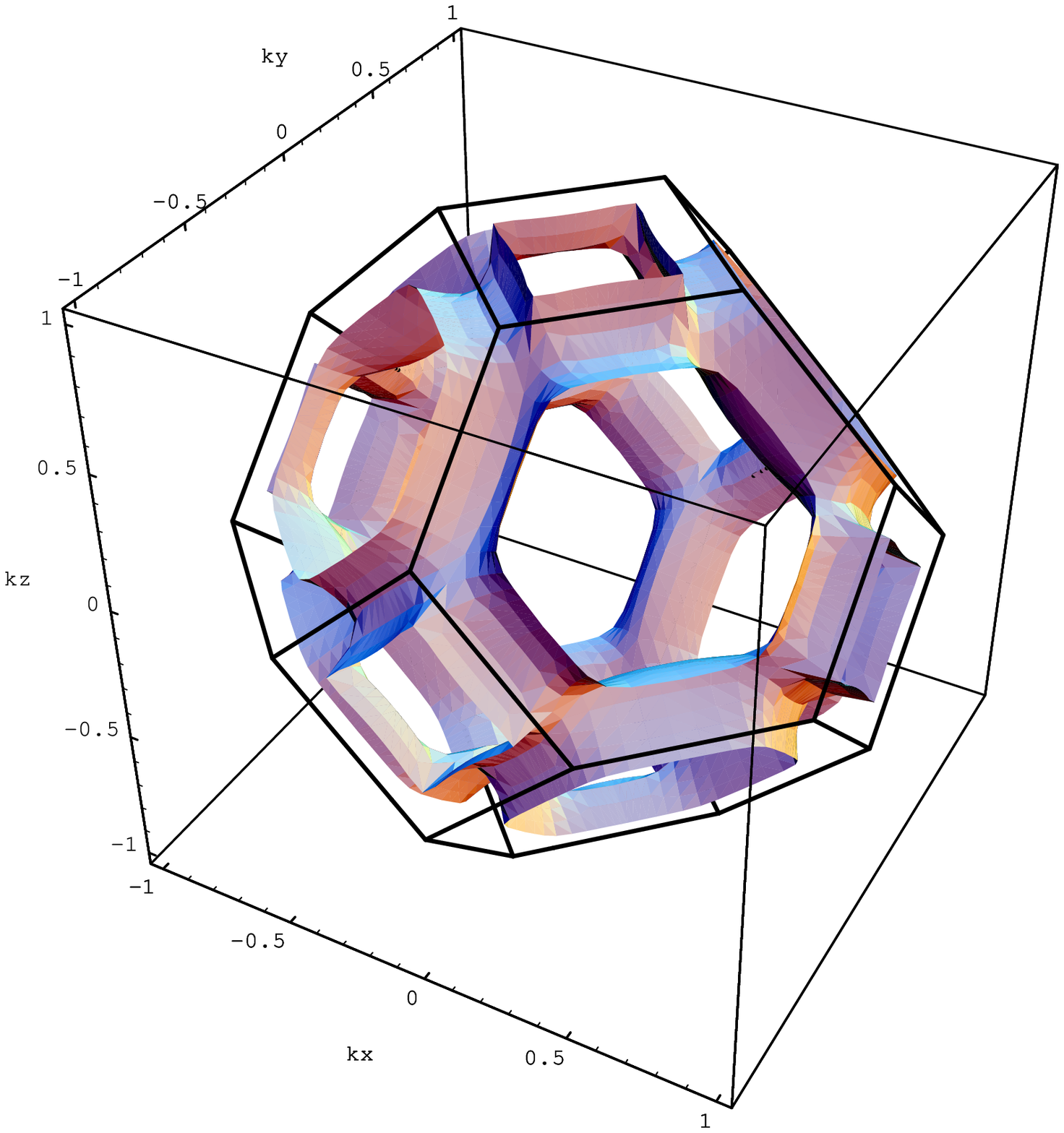,width=\figurewidth}
  \caption{
    \label{fig_Pb_3rd_band}
    (Color online) Third band in Pb calculated using the matrix by
    \citet{Anderson65}. The more complicated Fermi surface structure is
    illustrated by the ``pipe'' like tubes along the zone edges.
  }
\end{figure}

\citet{Anderson65} carried out de Haas--van Alphen measurements on Pb and their
subsequent analysis is based on the orthogonalized plane wave method which is
fit to de Haas--van Alphen data of the Fermi surface.  The Fermi energy,
$\epsilon_{F}$, the Fourier components of the pseudopotential parameters,
$V_{111}$and  $V_{200}$, and a spin--orbit coupling parameter $\lambda$ are
regarded as fitting parameters. We used the following values: $\epsilon_{F} =
9.765$ eV, $V_{111} = -1.142$ eV, $V_{200} = -0.530$ eV, and $\lambda = 1.306$
eV. Anderson and Gold's parameterization yields the dispersion relations for the
4 electron bands.  The Fermi surface cuts through the second and third bands,
which are shown in Figs.~\ref{fig_Pb_2nd_band} and \ref{fig_Pb_3rd_band}.
Comparing these figures with the form of the Fermi surface calculated for an
empty face center cubic lattice\cite{Ashcroft76:Solid_State_Physics}, one sees
that the effect of interactions among the electrons is to remove the sharp edges
in the second band and increase the density of electrons in the third band.
Electron interactions also now ensure that the fourth band is empty at low
temperatures whereas there is a small contribution to the electron density from
the fourth band in the empty lattice calculation.  By using this experimentally
determined Fermi surface we have included the effects of electron-electron
scattering in the dispersion of the second and third band carriers.

In order to capture the strong-coupling features, the contributions from
scattering among the different components of the Fermi surface have to be
evaluated very accurately.  These contributions are given by six-dimensional
integral and require a large number of mesh points.  The Anderson and Gold model
requires the diagonalization of a complex valued matrix which becomes
computationally prohibitive inside a numerical integration.  Fortunately,
superconductivity is effected by phonon exchange between electron states which
are close to the Fermi energy and we can approximate the dispersion by a power
series expansion about $k_{F}$ at each point on the Fermi surface and retain
only the linear term.  This is because the maximum phonon energy is much less
than the characteristic electronic energy scale and because there are no strong
features in the density of states close to the Fermi energy.  We therefore model
the electronic dispersion in the second and third bands using a simpler
relation,

\begin{equation}
  \label{eq_electron_dispersion}
  \xi_{\vec{k}}^{(i)} = \hbar v_{F}^{(i)}
    \left |  \left (\vec{k}  - \vec k_{F} \right)_{\perp} \right|,
\end{equation}

\noindent
where $v_{F}^{(i)}$ is the Fermi velocity on band $i$ as input and $ \left |
\left (\vec{k}  - \vec k_{F} \right)_{\perp} \right|$ is the magnitude of the
component of $\vec k - \vec k_{F}$ perpendicular to the Fermi surface at $\vec
k$.

\noindent
Using the Anderson and Gold matrix and averaging over the Fermi surface we find
a value of $v_{F}^{(2)} = -1.367 \times 10^{6} \,\,\, \mbox{m s${}^{-1}$} < 0$
and $v_{F}^{(3)} = 1.055 \times 10^{6} \,\,\, \mbox{m s${}^{-1}$}$. The second
band is hole--like and the third band electron--like. The location of the Fermi
surface in $k$--space is somewhat complicated but we can make some simplifying
assumptions. The second band is approximated well with a sphere and we use a
constant $k_{F}$ for $\xi_{\vec{k}}^{(2)}$.  The third band is more complicated
as shown in Fig.~\ref{fig_Pb_3rd_band}. The Fermi surface consists of ``pipes''
around the edges of the Brillouin zone. Details of the Fermi surface
parameterization used in the calculations are given in the Appendix, section
\ref{sec_Appendix}.

\section{Electron-Phonon Model}
\label{sec_electron_phonon_model}

Use of the experimental data ensures that the important consequences of the
electron-electron interactions at the low frequencies of interest here are
included in the location of the Fermi surface and in the Fermi velocities for
the second and third band carriers.  As a result the Hamiltonian contains a pure
band structure part, $\mathcal{H}_{band}$, and the electron-phonon Hamiltonian,
$\mathcal{H}_{el-ph}$.  The Hamiltonian is given by

\begin{equation}
  \mathcal{H} = \mathcal{H}_{band} + \mathcal{H}_{el-ph}
  = \sum_{\vec{k}\,\, \nu=2,3\,\, \sigma} \xi_{\vec{k} \, \nu} \,\,
    c^\dagger_{\vec{k} \nu \sigma} c_{\vec{k} \nu \sigma}
   \mbox{}
    + \sum_{\vec{k} \, \nu \,\nu' \,\sigma} \sum_{\vec{q} \vec{Q} \lambda}
    g_{\nu \,\nu'} ( \vec{q}, \, \vec{Q}, \, \lambda )
    ( a_{\vec{q} \lambda} + a^\dagger_{\vec{q} \lambda} )
    c^\dagger_{\vec{k} + \vec{q} + \vec{Q} \,\nu \,\sigma} c_{\vec{k} \nu' \,\sigma}.
\end{equation}

\noindent
In $\mathcal{H}_{band}$, the electron energies, $\xi_{\vec{k}\, \nu}$, are given
by our parameterization of the Anderson and Gold model eq.
(\ref{eq_electron_dispersion}), $\sigma$ denotes the electron spin, $\nu$
denotes the electronic band.  In $\mathcal{H}_{el-ph}$ the $a_{\vec{q} \lambda}$
and $a^\dagger_{\vec{q} \lambda}$ are the usual creation and destruction
operators for phonons and $\lambda$ is the branch index for the phonon spectrum.
The electron-phonon matrix element is

\begin{equation}
\label{ep_g}
g_{(\nu \nu')} (\vec{q}, \vec{Q}, \lambda) =
  -
  i \sqrt{ \frac{ \hbar^{2} }{ 2 N_{0} \, M_{\mathrm{Pb}} \, \hbar \omega_{\vec{q}
\lambda} } } \,\,\,
  \hat{\eta}_{\vec{q} \lambda} \cdot
  \left( \vec{q} + \vec{Q} \right)
  V_{(\nu \nu')} (\vec{q} + \vec{Q}),
\end{equation}

\noindent
where the phonon spectrum, $\hbar \omega_{\vec{q} \lambda}$, and the phonon
wavefunctions, $\hat{\eta}_{\vec{q} \lambda}$, are given by a force constant
model for the phonon dispersion based on work done by \citet{Cowley74a}.
$V_{(\nu \nu')}(\vec{q})$  is the Fourier transform of the screened electron-ion
pseudopotential describing scattering between the two electron bands in the
problem. We modeled the three different electron-phonon matrix elements,
$g_{(22)}(\vec{q}, \vec{Q}, \lambda)$, $g_{(23)}(\vec{q}, \vec{Q}, \lambda)$,
$g_{(32)}(\vec{q}, \vec{Q}, \lambda)$ and $g_{(33)}(\vec{q}, \vec{Q}, \lambda)$,
by a common pseudopotential function\cite{Ashcroft68} and scaling factors, so
that $g_{(\nu \nu')}(\vec{q}, \vec{Q}, \lambda) = a_{(\nu \nu')} \,\, g
(\vec{q}, \vec{Q}, \lambda)$.  The values of the $a_{(\nu \nu')}$ are determined
by fitting the solution of the gap equation to the experimentally determined
values for the two bands\cite{Farnworth74}, 1.29~meV and 1.38~meV, and are found
to be $a_{(22)} = 0.61$ , $a_{(23)}=0.50$ , and $a_{(33)}=
0.56$\cite{Bocklater}.

\section{Optical Conductivity}
\label{sec_results_optical_conductivity}

The optical conductivity is given by

\begin{equation}
\label{opticalcon}
\sigma(\omega) = -i\left(\frac {e^2}{\omega}\right )\chi(\omega) +
i\frac {\omega^2_{pl}}{4\pi \omega},
\end{equation}

\noindent
where $\chi(\omega) = \left< \left< j;j \right> \right> = -i \int^{\infty}_0 dt
\left< \left[ j(t),j(0) \right] \right> e^{i \omega t}$ is the retarded
current-current correlation function.  In the normal state the d.c.
conductivity, $\sigma(0)$, is finite so that the poles in the two terms exactly
cancel.  Consequently $\chi(0) = \omega^2_{pl} / \left( 4 \pi e^2 \right) = n \,
e^2 / m^*_{opt}$, where $n$ is the density of carriers and $m^*_{opt}$ is the
optical effective mass.  G\"{o}tze and W\"{o}lfle reformulated this result in
terms of a memory function, $M(\omega)$,

\begin{equation}
  \sigma(\omega) = \frac {i}{4\pi}
    \frac {\omega^2_{pl}}{ \omega + M(\omega )},
\end{equation}

\noindent
where

\begin{equation}
  \label{memoryfunction}
  M(\omega) = \frac {\omega \chi(\omega)}{\chi(0) - \chi(\omega)}.
\end{equation}

\noindent
In this formulation the cancellation of the poles at $\omega=0$ is accomplished
by introducing the memory function, $M(\omega)$, which  remains to be
determined.  The analytic properties of the correlation functions require that
the real part of $M(\omega)$, $M^{\prime}(\omega)$, is odd in $\omega$ and that
the imaginary part, $M^{\prime\prime}(\omega)$, is even.  There are several
potential contributions to $M(\omega)$ from different scattering mechanisms
depending on the system being considered.  G\"{o}tze and W\"{o}lfle considered a
number of mechanisms including impurity scattering and electron-phonon
scattering, among others\cite{Goetze72}.

\subsection{Electron-Phonon Contribution}

Here we are interested in the contribution from electron-phonon scattering which
is characterized by the strength of the coupling, $g$.  The functional form of
$M(\omega )$ is found from $\chi (\omega)$ to second order in $g$ by assuming
that $M(\omega)$ is a power series in the strength of the electron-phonon
interaction, $g$, so that, expanding the right hand side of eq.
(\ref{memoryfunction}) one finds\cite{Goetze72}

\begin{equation}
  M(\omega) = \frac { \left[ \left< \left< \left[ j, H \right];
    \left[ j, H \right] \right> \right>_{\omega}
    - \left< \left< \left[ j, H \right];
    \left[ j, H \right] \right> \right>_{\omega = 0} \right]}{\omega \chi(0)}
  + O(g^4),
\end{equation}

\noindent
where

\begin{equation}
  [j_i, H] = \sum_{\vec k, \nu, \vec k^{\prime}, \nu^{\prime}, \sigma}
\left[ \frac {\partial \xi_{\vec k, \sigma, \nu}}{\partial k_i}
      -\frac {\partial \xi_{\vec k^{\prime}, \sigma, \nu^{\prime}}}{\partial
k^{\prime}_i} \right ]
c^{\dagger}_{\vec k, \sigma, \nu}c_{\vec k{^\prime}, \sigma, \nu^{\prime}}
(a^{\dagger}_{\vec k -\vec k^{\prime}, \lambda} + a_{-\vec k +\vec k^{\prime},
\lambda}),
\end{equation}

\noindent
and

\begin{equation}
j_i = \sum_{\vec k, \sigma, \nu}
  \frac{\partial \xi_{\vec k, \sigma, \nu}}{\partial k_i}
  c^{\dagger}_{\vec k, \sigma, \nu} c_{\vec k, \sigma, \nu}.
\end{equation}

In the  superconducting state the response of the supercurrent to the external
field leads to a delta-function in $\sigma(\omega)$ at $\omega = 0$ in the real
part  and a pole in the imaginary part.  The weight of the pole is given by $c^2
/ \left( 4 \pi \lambda^2 \right)$ (see
\citet{Tinkham80:Introduction_to_Superconductivity}), where $\lambda$ is the
penetration depth.  Using experimental values for Pb, $\chi(0)$ in the
superconducting state is reduced from the normal state value by $\simeq 0.026$.
This leads to a negligible quantitative effect for the frequencies of interest
here, $\omega > 2 \Delta $, and therefore we will ignore this.

In the following we drop the band index to simplify the notation.  The
correlation function in the superconducting state to lowest order in the
strength of the electron-phonon interaction, $O(g^2)$, is

\begin{equation}
  \left< \left< \left[ j, H \right];
    \left[ j, H \right] \right> \right>_{\omega} =
    \left< \left< \left[ j, H \right];
    \left[ j, H \right] \right> \right>^{(N)}_{\omega}
    + \left< \left< \left[ j, H \right];
    \left[ j, H \right] \right> \right>^{(S)}_{\omega},
\end{equation}

\noindent
where

\begin{eqnarray}
  \lefteqn{
    \left< \left< \left[ j, H \right];
      \left[ j, H \right] \right> \right>^{(N)}_{\omega} =
    \sum_{\vec k, \vec q, \lambda}
    \left[ u^2_{\vec k} u^2_{\vec k + \vec q} + v^2_{\vec k} v^2_{\vec k + \vec q} 
    -2u_{\vec k} v_{\vec k} u_{\vec k + \vec q} v_{\vec k + \vec q} \right]
    \left| {\cal M}(\vec k, \vec q, \lambda) \right|^2
  }
\nonumber \cr
& & \mbox{} \times \left[
  f_{\vec k + \vec q}(1 - f_{\vec k})(1 + n_{\vec q}) -f_{\vec k }(1 - f_{\vec k
  + \vec q})n_{\vec q}
  \right]
  \left( \frac {1}{ E_{\vec k} + \omega_{\vec q} - E_{\vec k +\vec q} - \omega }
  + \frac {1} { E_{\vec k} + \omega_{\vec q} - E_{\vec k +\vec q} + \omega }
  \right)
\end{eqnarray}

\noindent
and

\begin{eqnarray}
  \lefteqn{
    \left< \left< \left[ j, H \right];
      \left[ j, H \right] \right> \right>^{(S)}_{\omega} =
    \sum_{\vec k, \vec q, \lambda}
    \left[ u^2_{\vec k} v^2_{\vec k + \vec q} + v^2_{\vec k} u^2_{\vec k + \vec q} +2
    u_{\vec k} v_{\vec k} u_{\vec k + \vec q} v_{\vec k + \vec q} \right ]
    \left| {\cal M}(\vec k, \vec q, \lambda) \right|^2
  }
\nonumber \cr
& & \mbox{} \times \left[
(1 - f_{\vec k + \vec q})(1 - f_{\vec k})\left ( 
\frac {2(\omega_{\vec q} + E_{\vec k +\vec q} + E_{\vec k})(1 + n_{\vec q})}
{(\omega_{\vec q} + E_{\vec k +\vec q} + E_{\vec k})^2 - \omega^2}
+\frac {2(\omega_{\vec q} - E_{\vec k +\vec q}- E_{\vec k})n_{\vec q}}
{(\omega_{\vec q} - E_{\vec k +\vec q}- E_{\vec k})^2 - \omega^2} \right )
\right.
\nonumber \cr
& & \mbox{} -
\left.
f_{\vec k } f_{\vec k + \vec q}\left ( 
\frac {2(\omega_{\vec q} - E_{\vec k +\vec q}- E_{\vec k})(1 + n_{\vec q})}
{(\omega_{\vec q} - E_{\vec k +\vec q}- E_{\vec k})^2 -\omega^2}
+\frac {2(\omega_{\vec q} + E_{\vec k +\vec q}+ E_{\vec k})n_{\vec q}}
{(\omega_{\vec q} + E_{\vec k +\vec q}+ E_{\vec k})^2  -\omega^2} \right)
 \right].
\end{eqnarray}

\noindent
The $u_{\vec k}$ and $v_{\vec k}$ are the usual BCS coherence factors.  ${\cal
M}(\vec k, \vec q, \lambda)$ is the electron-phonon interaction defined earlier
together with the derivatives of the quasiparticle spectrum in the current
operator,

\begin{equation}
{\cal M}(\vec k, \vec q, \lambda) =
-i \sqrt{ \frac {N_0\hbar^2}{2M_{\mathrm{Pb}}\hbar \omega_{\vec k -  \vec
k^\prime}^{\lambda}}}  
\left [\vec k -\vec k^\prime\cdot \eta^{\lambda}_{\vec k - \vec k^\prime} \right
] {V_{(\nu \nu')}}(\vec k - \vec k^\prime)
\left[ \frac {\partial \xi_{\vec k, \sigma, \nu}}{\partial k_i}
      -\frac {\partial \xi_{\vec k^{\prime}, \sigma, \nu^{\prime}}}{\partial
k^{\prime}_i} \right ]
\end{equation}

\noindent
$\left< \left< \left[ j, H \right];\left[ j, H \right] \right>
\right>^{(N)}_{\omega}$ has the character of a particle-hole contribution  and
smoothly evolves into the normal state result derived by \citet{Allen71} and by
\citet{Goetze72}.  It is exponentially small for temperatures much smaller than
the gap and is negligible at the temperature at which Farnworth and Timusk took
their data on lead in the superconducting phase, $0.35$ K.  On the other hand
the combination of coherence factors and the particle-particle or hole-hole
distribution functions ensures that $\left< \left< \left[ j, H \right];\left[ j,
H \right] \right> \right>^{(S)}_{\omega}$ vanishes as the gap vanishes.  This
term gives the conductivity at finite frequencies at low temperatures much less
than $T_{c}$.  Taking the limit as $T \rightarrow 0$, the real and imaginary
parts of $M_S(\omega)$ become

\begin{eqnarray}
M^{S\prime}(\omega) &=&
\sum_{\vec k, \vec q}
\left [ u^2_{\vec k} v^2_{\vec k + \vec q} + v^2_{\vec k} u^2_{\vec k + \vec q} +2
u_{\vec k} v_{\vec k} u_{\vec k + \vec q} v_{\vec k + \vec q} \right ]
\frac {|{\cal M}(\vec k, \vec q, \lambda)|^2}{\chi(0)}
\cr
& & \mbox{} \times
\frac { 2\omega}{ [(\omega_{\vec q} + E_{\vec k +\vec q} + E_{\vec
k})^2 - \omega^2][\omega_{\vec q} + E_{\vec k +\vec q} + E_{\vec k}] }
\cr
\cr
M^{S\prime \prime}(\omega) &=&
\pi \sum_{\vec k, \vec q}
\left [ u^2_{\vec k} v^2_{\vec k + \vec q} + v^2_{\vec k} u^2_{\vec k + \vec q} +2
u_{\vec k} v_{\vec k} u_{\vec k + \vec q} v_{\vec k + \vec q} \right ]
\frac {|{\cal M}(\vec k, \vec q, \lambda)|^2}{\chi(0)}
\cr
& & \mbox{} \times \left ( \frac {\delta [\omega -(\omega_{\vec q} + E_{\vec k +\vec q}+
E_{\vec k})]
+ \delta [\omega +(\omega_{\vec q} + E_{\vec k +\vec q}+ E_{\vec k})]
}{\omega_{\vec q} + E_{\vec k +\vec q}+ E_{\vec k}} \right )
\end{eqnarray}

\subsection{Absorptivity}
\label{sec_absorptivity}

The absorptivity is proportional to the real part of the surface impedance,
$\left (  1 + \left( 4 \pi i / \omega \right) \sigma (\omega) \right )^{-1/2}$.
As a minimal model we ignore the difference between the values of the gaps on
the Fermi surfaces associated with the second and third bands and take
$\Delta_{0} = 1.37$~meV. This is an average of the two values found by
\citet{Farnworth74}.  We will therefore get three contributions to the memory
function, $M (\omega)$, and to the absorptivity.  These three are from
intra--band scattering within the second band, $M_{(22)} (\omega)$, intra--band
scattering within the third band, $M_{(33)} (\omega)$, and inter--band
scattering between the second and the third band, $M_{(23)} (\omega)$. Since the
frequencies of interest are much smaller than the plasma frequency,
$\omega_{pl}$, the expression for the absorptivity $A (\omega)$ can be greatly
simplified.

\begin{equation}
A (\omega)  \simeq 
  \frac{4\omega}{\omega_{pl}}
  \sqrt{ 2 \left ( \left | 1+ \frac{ M(\omega)} {\omega}\right |
- \left [1 + \frac{ M^{\prime } (\omega) }{\omega} \right ] \right )}
 \simeq
 \frac{2}{\omega_{pl}}
 \frac {M^{\prime \prime}(\omega)} 
{\sqrt{1 + \frac{ M^{\prime } (\omega)}{\omega}}} 
\end{equation}

\noindent
Here we have also assumed that $1 + \frac{ M^{\prime } (\omega) }{\omega} >1$
and $\frac {M^{\prime \prime}(\omega)}{\omega} < 1$ so that square root can be
expanded.  $\frac{ M^{\prime } (\omega) }{\omega} $ evaluated at $\omega =0$ is
the enhancement of the optical effective mass which is similar to the
enhancement of the effective mass due to the electron-phonon interaction in the
specific heat.

\begin{figure}
  \psfig{file=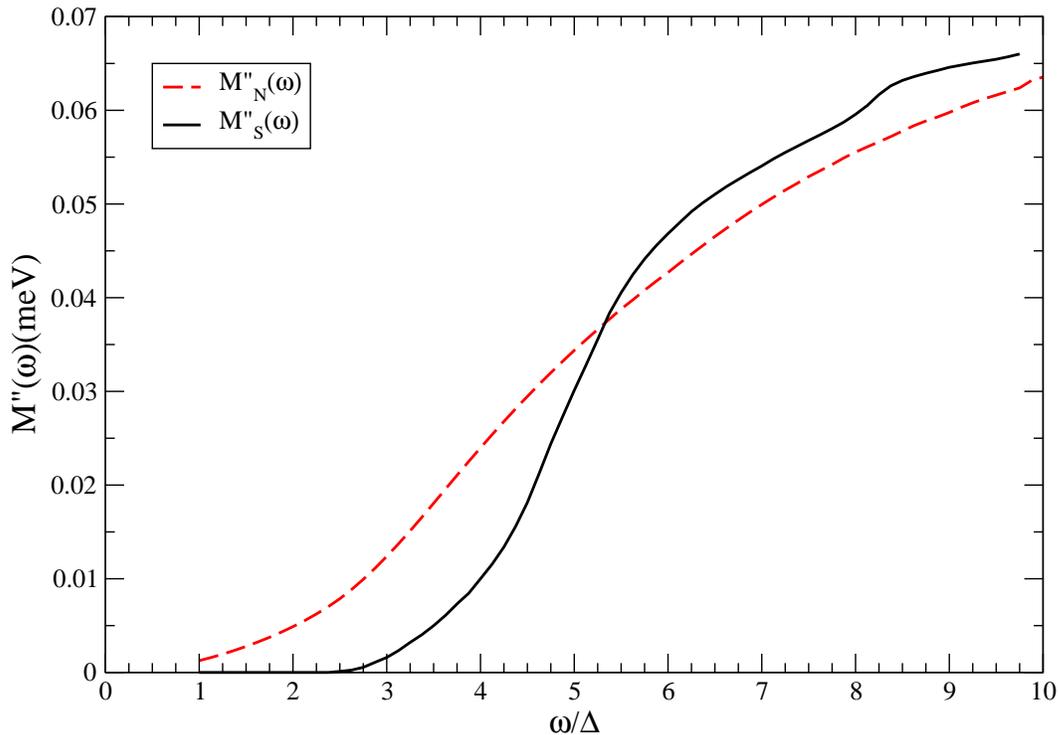,angle=-90,width=\figurewidth}
  \caption{
    \label{fig_OC22}
    (Color Online)   $ M^{\prime \prime}_{(22)} (\omega)$ vs. $\omega$. The full
    line is $ M^{S\prime \prime}_{(22)} (\omega)$ in the superconducting state
    while the dashed line is $ M^{N\prime \prime}_{(22)} (\omega)$ in the normal
    state.
  }
\end{figure}

\begin{figure}
  \psfig{file=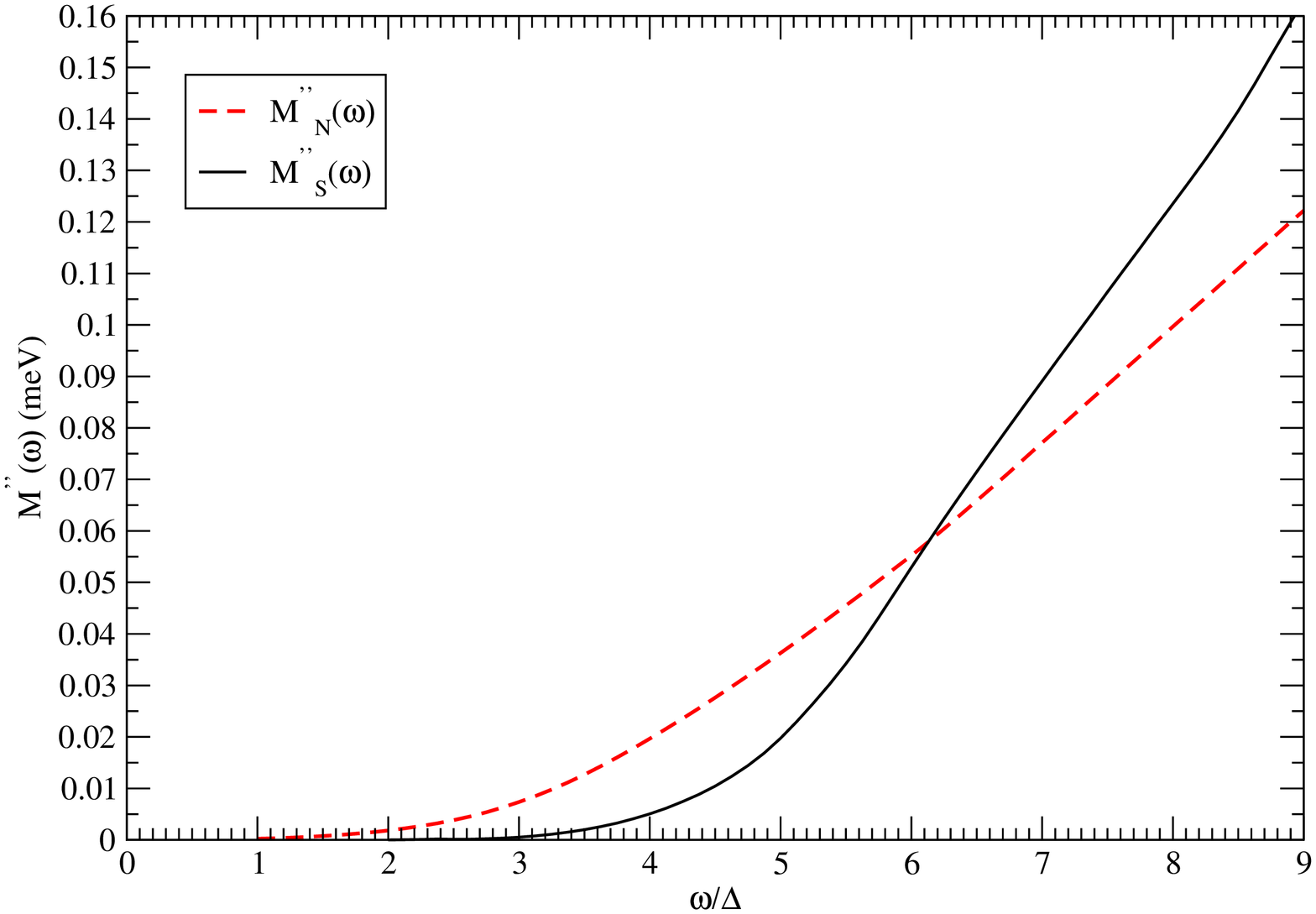,angle=-90,width=\figurewidth}
  \caption{
    \label{fig_OC23}
    (Color Online)   $ M^{\prime \prime}_{(23)} (\omega)$ vs. $\omega$. The full
    line is $ M^{S\prime \prime}_{(23)} (\omega)$ in the superconducting state
    while the dashed line is $ M^{N\prime \prime}_{(23)} (\omega)$ in the normal
    state.
  }
\end{figure}

\begin{figure}
  \psfig{file=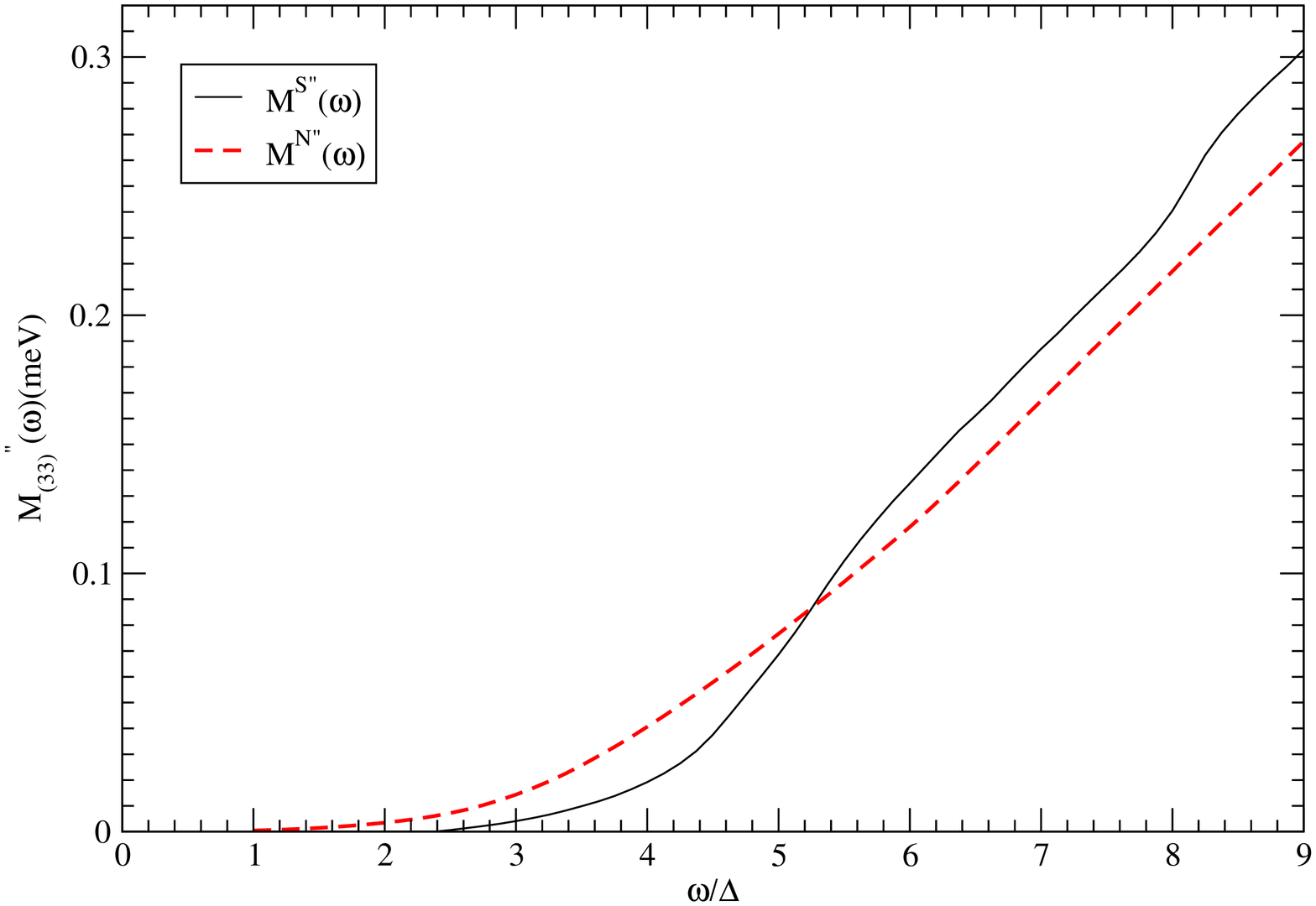,angle=-90,width=\figurewidth}
  \caption{
    \label{fig_OC33}
    (Color Online)  $M^{\prime \prime}_{(33)} (\omega) $ vs. $\omega$. The
    dashed line is $M^{S\prime \prime}_{(33)} (\omega)$ in the superconducting
    state while the full line is $ M^{N\prime \prime}_{(33)} (\omega) $ in the
    normal state.
  }
\end{figure}

The $ M^{\prime \prime}_{(22)} $, $M^{\prime \prime}_{(23)} (\omega)$, and $
M^{\prime \prime}_{(33)} $ contributions in the normal and superconducting
states to order $g^{2}$ are compared in Figs.~\ref{fig_OC22}, \ref{fig_OC23},
and \ref{fig_OC33}.  The different magnitudes of the memory functions $M_{(ij)}
(\omega)$, reflects their dependence on the surface area of the different parts
of the Fermi surface involved in the scattering processes and on the size of the
phase space for phonons which connect different parts of the Fermi surface.
This dominates any effects the density of states on the different Fermi surfaces
might have due to the cancellation of the Fermi velocity factors in the density
of states with those from the current operator matrix elements.

The dominant contribution comes from $ M^{\prime \prime}_{(33)} (\omega) $ which
involves the scattering between the 36 ``pipes'' which make up the third band.
Although the geometry of the third band suggests that quasi-one dimensional
``nesting'', scattering between parallel pipes with a single phonon momentum,
could be important, it is found not to dominate $ M^{\prime \prime}_{(33)}
(\omega) $. There are $\frac {36^2}{2} = 648$ different combinations of pipes
and of those, pipes whose mid--points can be connected by momentum vectors
$\vec{Q}_{1} = 2 \pi / a \left( 1, 1/2, 1/2 \right)$ and $\vec{Q}_{2} = 2 \pi /
a \left( 3/2, 0, 0 \right)$ lead to contributions much larger than other
combinations by about a factor of $\sim 10$. There are only approximately 100 of
these combinations and their contribution does not dominate $ M^{\prime
\prime}_{(33)} (\omega) $. 

In contrast to the intra--band scattering events on the second band, there are
many more phonons that connect the more complicated surface of the third band.
This and the larger surface area of the third zone Fermi surface is responsible
for the big difference in magnitude of the two contributions.

The real parts of the $M_{(ij)}$ are related to the imaginary parts through a
Kramers--Kronig relation, $M^{\prime}(\omega) =
\frac{2\omega}{\pi}\int^{\infty}_0dz \frac{M^{\prime\prime}(z)}{z^2 -\omega^2}$,
so that $M^{\prime}(\omega)$ can, in principle,  be calculated directly from
$M^{\prime\prime}(\omega)$.  Consequently the magnitudes of the real parts of
the $M_{(ij)}$ have a similar relation to one another as the imaginary parts.
$\lim_{\omega \rightarrow 0}M^{\prime } (\omega) / \omega$ corresponds to the
optical effective mass enhancement due to the electron--phonon interaction,
similar to the enhancement seen in the specific heat.  $M^{\prime} (\omega) /
\omega$ in the superconducting and normal states are smooth functions of
$\omega$ which initially increase from their $\omega = 0$ values and reach
maximum values at approximately $3 \Delta$ in the normal state and approximately
$5 \Delta$ in the superconducting state before falling off monotonically with
increasing $\omega$.  Beyond $\omega = 6 \Delta$ they are almost equal in
magnitude.

\subsection{Strong-Coupling Features}
\label{sec_strong-coupling}

In order to analyze the results further, we follow Farnworth and Timusk and take
the derivative of $A (\omega)$ with respect to $\omega$. This reveals the
strong--coupling electron--phonon features.

\begin{equation}
  \frac{ d A_{S} (\omega) }{ d\omega }
    - \frac{ d A_{N} (\omega) }{ d\omega } \propto
\frac{ d M^{S\prime\prime} (\omega) }{ d\omega }
    - \frac{ d M^{N\prime\prime} (\omega) }{ d\omega }
\end{equation}

\begin{figure}
 \psfig{file=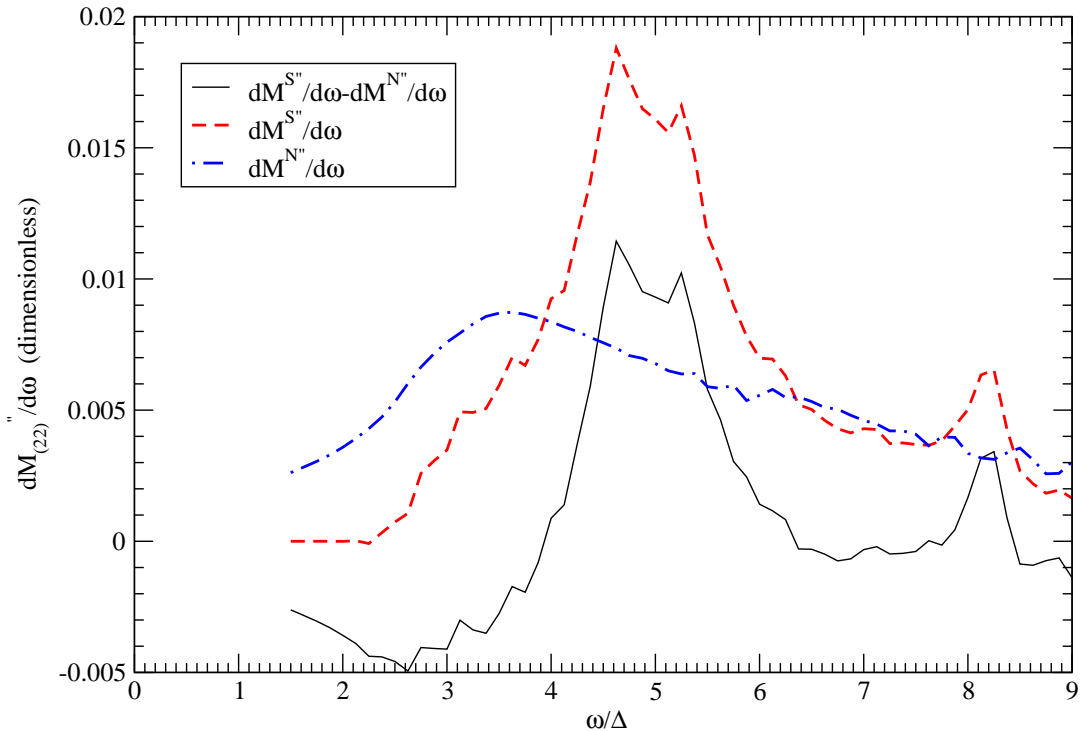,angle=-90,width=\figurewidth}
  \caption{
    \label{fig_OC22_der}
    (Color Online) The difference between $d^{\prime \prime}M_{(22)} (\omega) /
    d\omega$ (full line) in the superconducting and normal states.
  }
\end{figure}

\begin{figure}
 \psfig{file=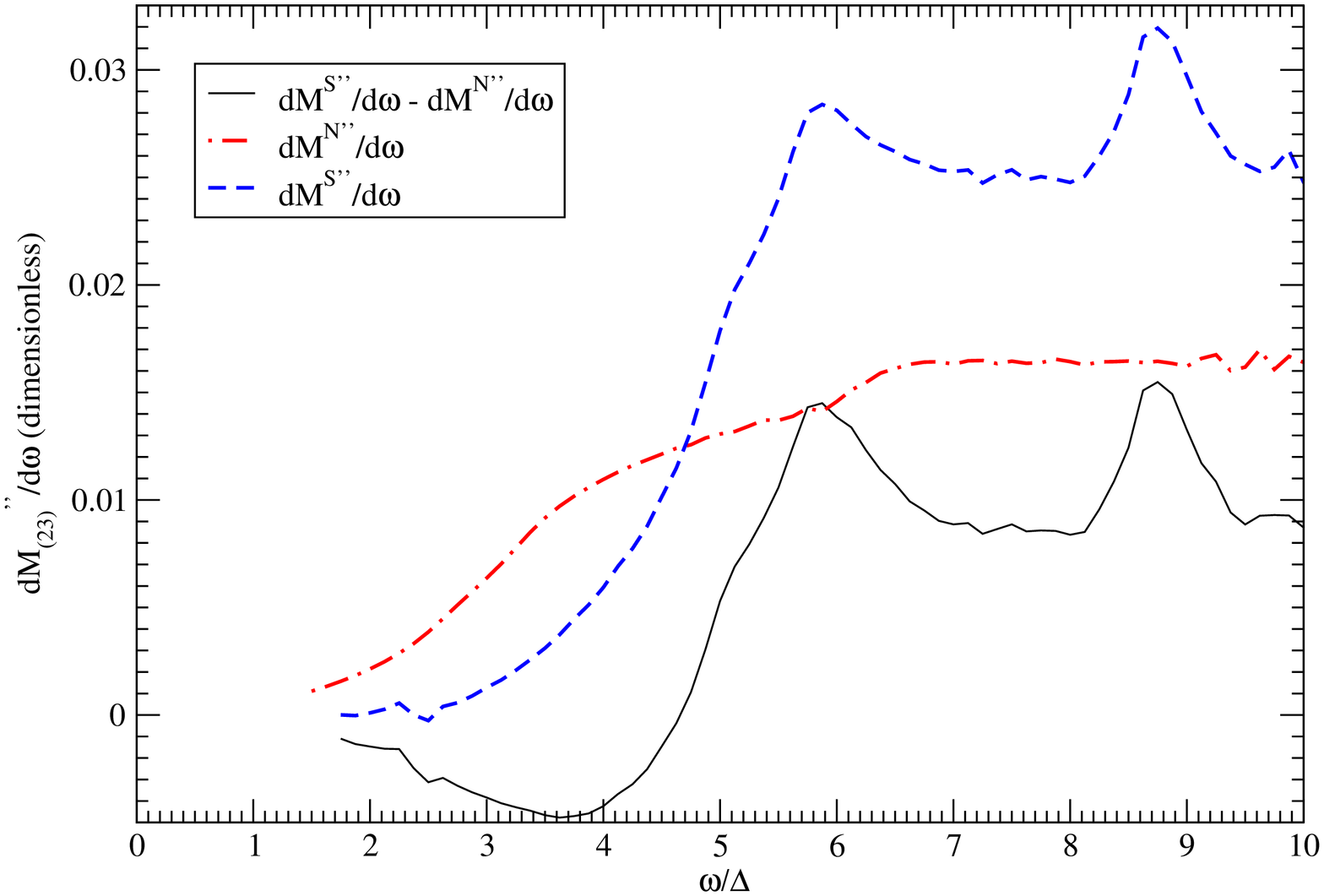,angle=-90,width=\figurewidth}
  \caption{
    \label{fig_OC23_der}
    (Color Online) The difference between $d^{\prime \prime}M_{(23)} (\omega) /
    d\omega$ (full line) in the superconducting and normal states.
  }
\end{figure}

\begin{figure}
  \psfig{file=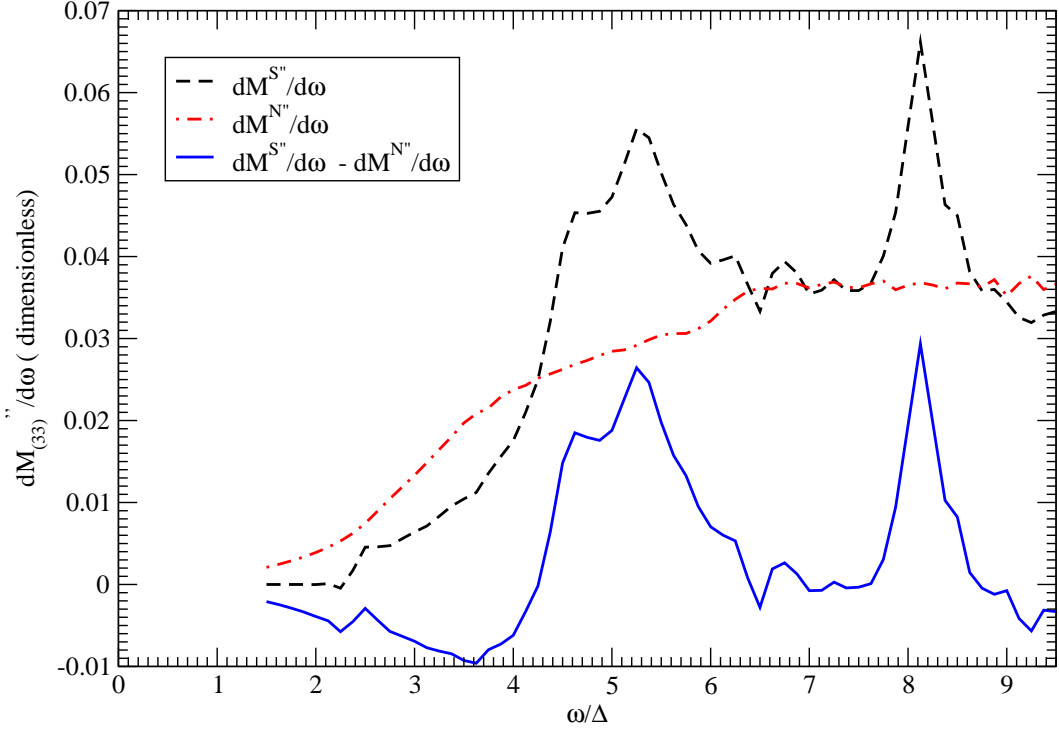,angle=-90,width=\figurewidth}
  \caption{
    \label{fig_OC33_der}
    (Color Online) The difference between $d^{\prime \prime}M_{(33)} (\omega) /
    d\omega $ (full line) in the superconducting and normal states. Note that
    the peak at $\sim 8.5$~meV has the same height as that at 5.5~meV. This
    reproduces an important feature of the Farnworth and Timusk
    data\cite{Farnworth76}.
  }
\end{figure}

\begin{figure}
  \psfig{file=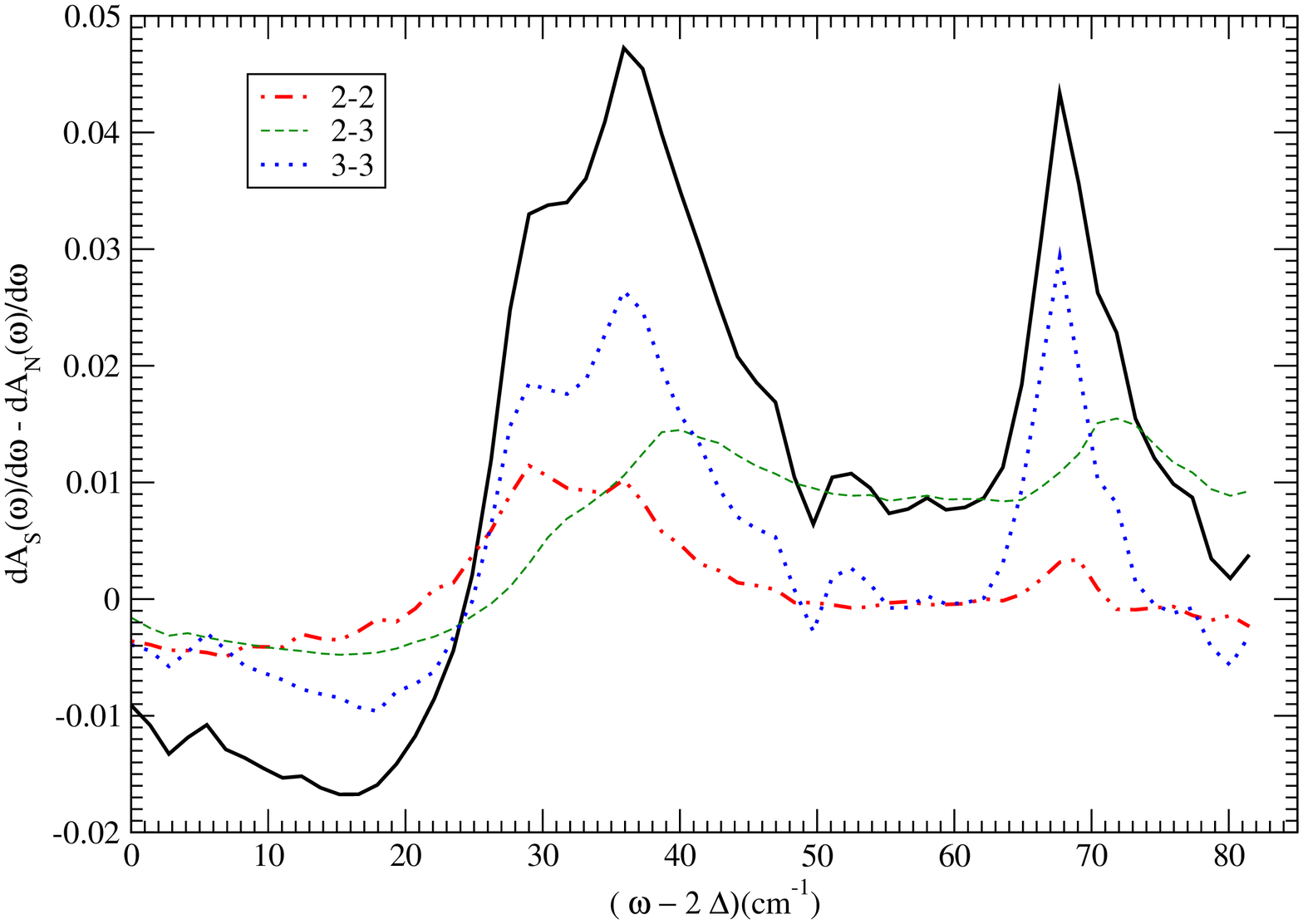,angle=-90,width=\figurewidth}
  \caption{
    \label{fig_contributions}
    (Color Online) Contributions to $d A_{S} (\omega) / d\omega - d A_{N}
    (\omega) / d\omega $ plotted against $\omega -2 \Delta$ in units of
    cm${}^{-1}$(full line) from the three types of scattering processes.
  }
\end{figure}

The strong-coupling features come almost exclusively from the first term on the
right-hand side of the equation.  In figures \ref{fig_OC22_der},
\ref{fig_OC23_der}, \ref{fig_OC33_der}, and \ref{fig_contributions} we compare
the contributions from the different scattering processes to $d
M^{S\prime\prime} (\omega) / d\omega$.  In each case the peaks at $\simeq
35$~cm${}^{-1}$ and $\simeq 70$~cm${}^{-1}$, present in the data of Farnworth
and Timusk, are recovered.  These peaks come from features in the phonon density
of states and are associated with a wide range of wavenumbers.  The $M^{N \prime
\prime}_{(ij)} (\omega)$ in the normal state is featureless above $4 \Delta$ and
subtracting from the superconducting $M^{S \prime \prime}_{(ij)} (\omega)$ does
not introduce features not already present.

$M^{S\prime \prime}_{(33)} (\omega) $ is the dominant contribution to $
M^{S\prime \prime} (\omega) $ and its derivative strongly resembles the data in
\citet{Farnworth76}.  The two prominent peaks at C and G closely resemble the
peaks at the same energies in Fig.~\ref{fig_OC33_der} from the point of view of
both height and width. The features at B, D and E are also seen in
Fig.~\ref{fig_OC33_der} and the relative magnitude of the calculated $dM^{\prime
\prime} / d\omega$  at E  and at C and G is the same as in the data.  The
interband scattering contribution, $dM^{\prime \prime}_{(23)} / d\omega$,
enhances equally the magnitudes of the peaks at C and G but shows none of the
other features.  These differences arise from the shape of the two contributions
to the Fermi surface from the two bands. Scattering between different parts of
the third band Fermi surface introduces all the phonon modes and as a result it
is the dominant contribution and closely resembles the phonon density of states.
By contrast scattering within the second band constrains the phase space for
phonons which conserve energy and momentum.  The magnitude of the higher energy
peak in $d M_{(22)}^{\prime\prime} (\omega) / d\omega$ is much smaller than that
of the low energy peak, whereas the peaks are of equal magnitude in $d
M_{(23)}^{\prime\prime}(\omega) / d\omega$ and $d
M_{(33)}^{\prime\prime}(\omega) / d\omega$.  The high energy peak comes from
phonons whose energy is $\sim 8.48$~meV.  In the case of $M_{22}(\omega)$ the
magnitude of the wavenumbers of the phonons with this energy lie between
1.33 \AA${}^{-1}$ and 1.43 \AA${}^{-1}$.  By contrast the corresponding phonon
wavenumbers in the case of $M_{33}(\omega)$ are grouped in sets with $|\vec q|$
$\sim 0.78$ \AA${}^{-1}$, $\sim 0.8$ \AA${}^{-1}$, $\sim 0.86$ \AA${}^{-1}$, and
$\sim 0.88$ \AA${}^{-1}$.  In Fig.~\ref{fig_contributions} we show the different
contributions to $dM^{\prime \prime} / d\omega$.

\begin{figure}
 \psfig{file=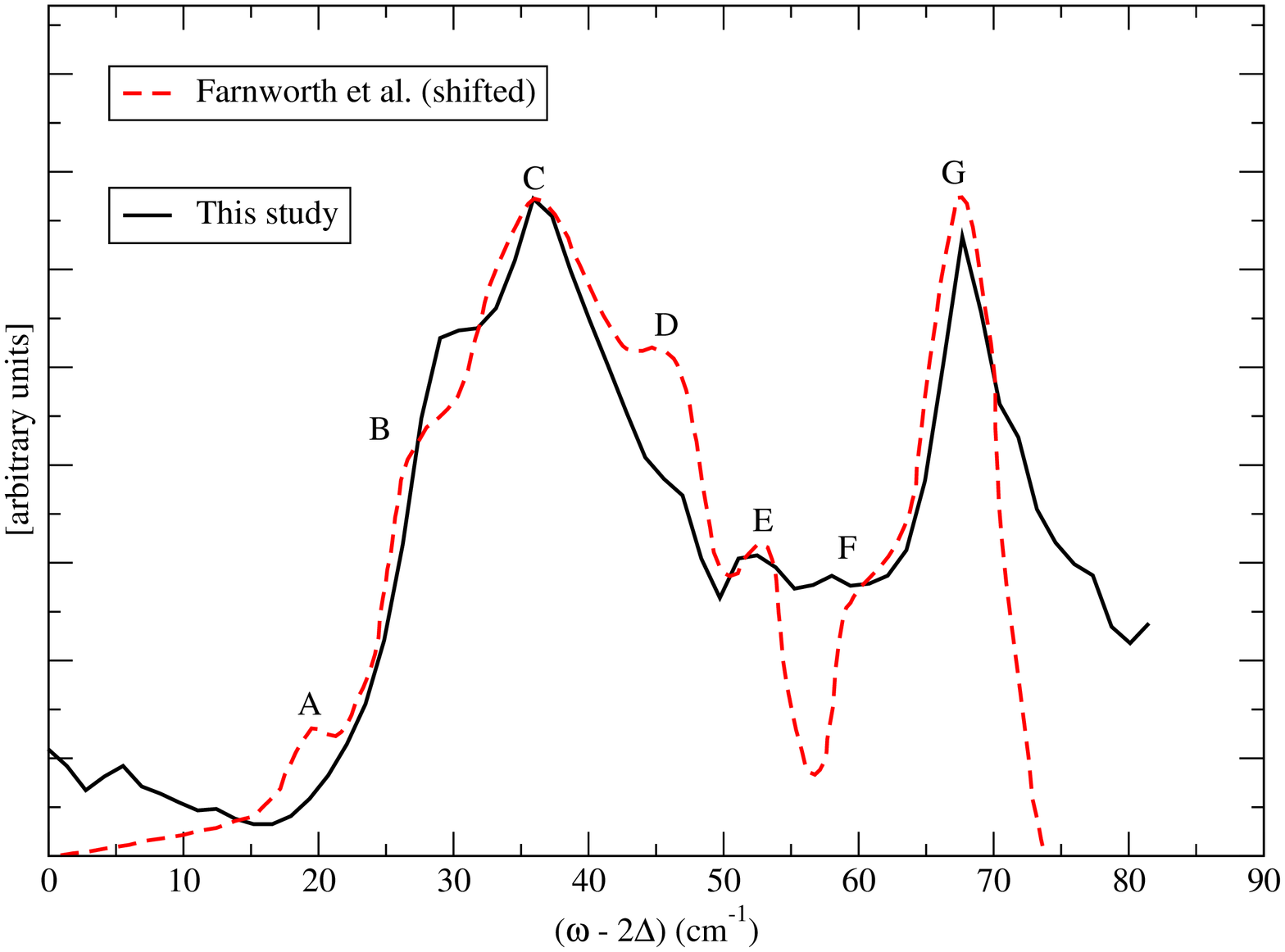,angle=-90,width=\figurewidth}
  \caption{
    \label{fig_timusk_comparison}
    (Color Online) $d A_{S} (\omega) / d\omega - d A_{N} (\omega) / d\omega $
    plotted against $\omega -2 \Delta$ in units of cm${}^{-1}$(full line). Many
    of the features of Farnworth and Timusk's data, shown in
    Fig.~\ref{fig_2_farnworth76} are reproduced.  Note that the peak G at $\sim
    8.5$~meV has the same height as the peak C at 5.5~meV.  This reproduces the
    most prominent feature of the Farnworth and Timusk data\cite{Farnworth76}.
  }
\end{figure}

In Fig.~\ref{fig_timusk_comparison} we plot the calculated difference of the
derivatives of $A_S(\omega)$ and $A_N(\omega)$ frequency shifted downwards by $2
\Delta$ for direct comparison with Farnworth and Timusk's data.  We have also
shifted Farnworth and Timusk's data by 0.9~cm${}^{-1}$ which takes account of
the slightly different values of $\Delta$ used by them and us.  Many of the
features of the experimental data are reproduced: the shoulder B ($\sim
27$~cm${}^{-1}$) is enhanced in our calculation, the peaks at C ($\sim
36$~cm${}^{-1}$) and at G ($\sim 67$~cm${}^{-1}$) are recovered and are equal in
magnitude as in the data, and the small feature at E ($\sim 52$~cm${}^{-1}$) is
reproduced.  The shoulder at D ($\sim 50$~cm${}^{-1}$) is much weaker in the
calculation compared to the data and the dip at $\sim 56$~cm${}^{-1}$ between E
and F is missing.  These discrepancies in the $dM^{S\prime \prime} / d\omega$
are probably due to our parameterization of the Fermi surface.  As we pointed
out earlier, the feature at A arises from contributions which are fourth order
in the electron-phonon coupling and so are naturally absent from our
calculation.

The contribution to $\sigma(\omega)$ from the Holstein process in Pb was
investigated many years ago by \citet{Allen71}.  He used the  isotropic
approximation familiar from the conventional Eliashberg formalism in which
momentum conservation is not enforced in the electron-phonon scattering.  In
this approach the details of the bandstructure are completely absent.  He
introduced a transport version of the $\alpha^{2} F (\omega)$, $\alpha_{tr}^{2}
F (\omega)$, in which the high frequency components were reduced to allow for
the suppression of forward scattering in the matrix element. He used this to
evaluated a frequency dependent relaxation time.  This amounts to putting in by
hand the strong-coupling features which we have calculated by taking the
derivatives of  the $M^{''}_{ij}(\omega)$'s.  Allen used the isotropic
approximation to calculate this contribution to $\sigma(\omega)$.  His result,
$\alpha_{tr}^{2} F (\omega)$ (Fig.~4 of \citet{Allen71}), is very similar to
$dM^{''}_{22}(\omega) / d\omega$ in Fig.~\ref{fig_OC22_der}.  The main features
at 4--5~meV and at $\sim 8$~meV are recovered.  However, as he remarked, there
is a discrepancy between the calculation of the absorptivity and the data of
Joyce and Richards\cite{Joyce70} and he offered a number of possible sources for
this.  Our calculation indicates that this discrepancy arises from the use of
the isotropic approximation for the Fermi surface.

Allen suggested that one source of the discrepancy between his calculation and
the data could be the absence of strong coupling corrections described by a
frequency dependent gap function as calculated from the Eliashberg equations.
Our calculation, using a momentum and frequency independent gap, reproduces much
of the fine structure seen in the strong-coupling features and it seems likely
that the differences with the data may be due to simplifications in the
parameterization of the Fermi surface or phonon spectrum.  However the absence
of structure in the gap function could lead to corrections.  Allen referred to a
formalism for including a gap with structure developed by \citet{Nam1967} which
incorporates details of the scattering mechanism by using the single-particle
self-energy determined by the Eliashberg equations. Apart from the lack of
momentum conservation in this conventional application of these equations,
which could be fixed by adding momentum dependence\cite{Choi2002a, Choi2002b},
the single-particle lifetime is not the same as the transport relaxation time
which is given by $M^{\prime\prime}(\omega)$ in our calculation.  A more
complete theory remains to be developed which incorporates both a realistic
Fermi surface and corrections to the constant gap approximation.

Finally, we comment on the identification of $M(\omega)$, an  ``optical
single-particle self-energy'', with the single-particle self-energy, as
extracted from ARPES by \citet{Hwang2004} in their analysis of data on
Bi$_2$Sr$_2$CaCu$_2$O$_{8+\delta}$.  In the electron-phonon model discussed here
the $M^{\prime \prime}_{(ij)} (\omega)$'s are quite different in form from the
well-known imaginary part of the single-particle self-energy due to the
electron-phonon interaction, $\Sigma^{\prime \prime}(E)$, as shown in
Figure~6.16 of \citet{Mahan90:Many--Particle_Physics}. Although $\Sigma(E)$ and
the $M (\omega)$ have the same analytic requirements, the real parts are odd in
frequency while imaginary parts are even, so that they have similar behavior at
low frequencies, their finite frequency behavior is quite different.  The
imaginary part of the self-energy is constant beyond $\sim 10$~meV or $\sim 7.3
\Delta_0$ whereas $ M^{\prime \prime}_{(ij)} (\omega) $'s grow monotonically as
the phase space for quasiparticle states increases with $\omega$.  Further, in
the superconducting state, a feature associated with a peak in the phonon
density of states, which appears at $\omega_{ph}$ (say), appears at
$\Delta+\omega_{ph}$ in $ \Sigma^{\prime \prime}(E) $ but at
$2\Delta+\omega_{ph}$ in $ M^{S\prime \prime} (\omega) $.  If further analysis
of data on Bi$_2$Sr$_2$CaCu$_2$O$_{8+\delta}$ continues to support the proposed
relation between $M(\omega)$ and $\Sigma(E)$, this will provide a strong
constraint on the underlying Hamiltonian.

\section{Conclusions}
\label{sec_conclusions}

Using the memory function formalism we have demonstrated the importance of
including a realistic Fermi surface in Pb to obtain quantitative agreement with
the strong-coupling features seen in the optical data.  The largest contribution
to the strong-coupling features comes from intra-third band scattering which
also  provides most of the fine structure.  The phonons involved in scattering
within this band are not as restricted by momentum conservation as those
involved in intra-second band scattering because the third band contribution to
the Fermi surface is spread over a substantial fraction of the region near the
zone boundary.  This seems to be the reason why the $\alpha^2F(\omega)$
determined from tunneling and that determined from $\sigma(\omega)$  are very
similar, whereas one might have assumed that suppression of forward scattering
in the current matrix elements would have lead to a difference between them.
Allen's investigation of the relation between $\alpha^2F(\omega)$ and
$\alpha_{tr}^2F(\omega)$ assumed a spherical Fermi surface and his model for
$\alpha_{tr}^2F(\omega)$ is very similar to the strong coupling features found
from scattering between states on a spherical Fermi surface, $dM^{\prime
\prime}_{(22)} / d\omega$ above.

Strong-coupling features in tunneling and transport data are a useful check on
microscopic models of the mechanism for superconductivity. There are a number of
systems in which mechanisms other than the electron-phonon interaction are
thought to play a role.  Among these are the cuprates where the contribution of
the electron-phonon interaction remains to be clarified and in which the
magnetic properties have been proposed as the origin of the d-wave
superconductivity. Analysis of data so far finds no evidence of electron-phonon
strong coupling features\cite{Basov2005}, although \citet{Maksimov2005} have
proposed a model for the electron-phonon interaction with which they have
analyzed ARPES data on Bi$_2$Sr$_2$CaCu$_2$O$_8$.  The data of \citet{Hwang2004}
show that $M^{\prime\prime}(\omega)$ for Bi$_2$Sr$_2$CaCu$_2$O$_{8+\delta}$ has
the same monotonic frequency dependence as was found in current work on Pb at
low temperatures so that its derivative may reveal strong-coupling features
associated with a magnetic mechanism.  Analysis of data on the cuprates to date
has employed the same formalism as in the electron-phonon problem with the
$\alpha_{tr}F(\omega)$ function replaced with a spectral density for spin
fluctuations.  The weak coupling approximation introduced by \citet{Goetze72}
would have to be extended to calculate the contributions to M($\omega)$ from
electron-phonon or impurity scattering enhanced by spin fluctuations or from
Umklapp scattering due to interactions among the charge carriers.
\citet{Riseborough83} has previously calculated the contribution from Umklapp
scattering due to spin fluctuations in the random phase approximation using the
Kubo formalism.  A memory function approach extended to deal with intermediate
to strong coupling could provide a method to test microscopic models for spin
fluctuation mechanisms for superconductivity in the cuprates while taking
account of bandstructure and momentum conservation.

\section{Acknowledgments}

We acknowledge P. Waltner for technical assistance with the representation of
the Fermi surface and   D. C. is grateful to L. Coffey for useful discussions.
D. C.  was supported in part by USDOE (DE-FG02-03ER46064).

\bibliography{paper-opt}
\bibliographystyle{apsrev}
\bibliographystyle{apsrev}
\newpage

\section{Appendix: Fermi Surface Parameterization}
\label{sec_Appendix}

Although diagonalizing the Anderson and Gold 8x8 matrix is computationally not
very challenging in itself, we will need to calculate its eigenvalues many times
since we will integrate over the electron momentum. It turns out that using a
diagonalization routine directly inside the integration is computationally not
feasible. We therefore had to simplify the model.

\begin{figure}
  \psfig{file=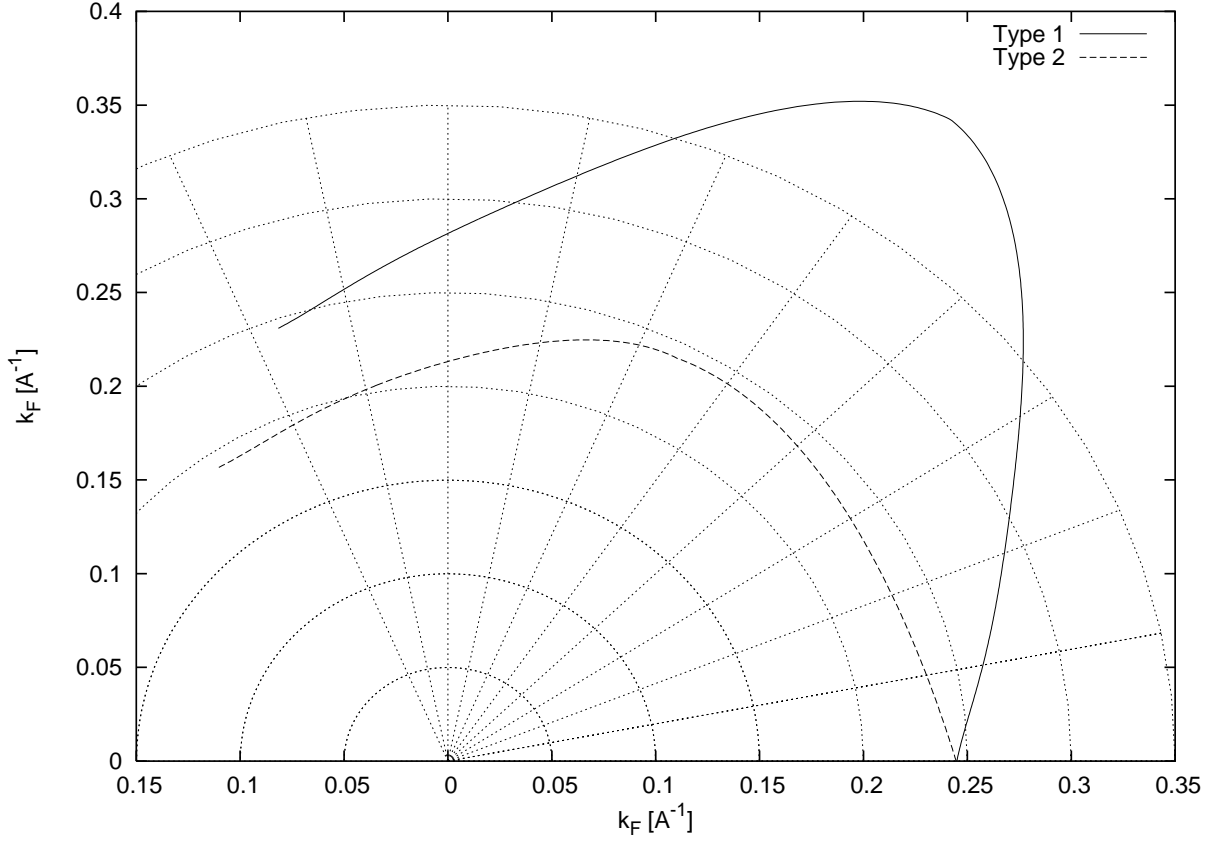,angle=-90,width=\figurewidth}
  \caption{
    \label{fig_FS_3rd_band}
    The Fermi momentum as a function of angle $\beta$ for the two types of third
    band pipes.
  }
\end{figure}

As pointed out earlier, we will require two of the four bands only. The second
band Fermi surface is close to spherical which led us to set the Fermi momentum
in that band to a constant, $v^{(2)}_{F} = -1.367 \times 10^{6}$ m s${}^{-1}$.
The second band is hole--like.

The third band Fermi surface consists of ``pipes'' which we approximated by
dented cylinders. There are two types of pipes and we chose to parametrize these
separately. In Fig.~\ref{fig_FS_3rd_band} we show the angular dependence of the
Fermi momentum on each of the two pipes. The dispersion was linearized and the
Fermi velocity we calculated for both types is $v^{(3)}_{F} = 1.055 \times
10^{6}$ m s${}^{-1}$. The third band is electron--like.

\end{document}